\RequirePackage{fix-cm}
\documentclass[twocolumn]{svjour3}          
\smartqed  
\usepackage{graphicx}
\journalname{Granular Matter}
\begin{document}


\title{Random packing of rods in small containers}

\author{Julian O. Freeman \and
Sean Peterson \and
Cong Cao \and
Yujie Wang \and
Scott V. Franklin \and
Eric R. Weeks}

\authorrunning{Freeman {\it et al.}} 

\institute{J.O.F., C.C., and E.R.W. \at
Department of Physics, Emory University, Atlanta, GA 30322, USA\\
\email{erweeks@emory.edu}
\and 
S.V.F and S.P. \at
School of Physics \& Astronomy, Rochester Institute of Technology, 
Rochester, NY, 14623, USA
\and
Y.W. \at 
Department of Physics, Shanghai Jiao Tong University, 
800 Dong Chuan Road, Shanghai 200240, China
}


\date{January 31, 2019}
\maketitle

\begin{abstract}
We conduct experiments and simulations to study the disordered packing of rods in small containers.  Experiments study cylindrical rods with aspect ratio ranging from 4 to 32; simulations use of spherocylinders with similar aspect ratios.  In all cases, rods pack randomly in cylindrical containers whose smallest dimension is larger than the rod length.  Packings in smaller containers have lower volume fractions than those in larger containers, demonstrating the influence of the boundaries.  The volume fraction extrapolated to infinite container size decreases with increasing aspect ratio, in agreement with previous work.  X-ray tomography experiments show that the boundary effects depend on the orientation of the boundary, indicating a strong influence of gravity, whereas the simulation finds boundary effects that are purely geometric.  In all cases, the boundary influence extends approximately half a particle length into the interior of the container. 
\end{abstract}

\section{Introduction}

People have long been curious about how objects pack into containers. Kepler conjectured that spheres
pack most efficiently when arranged into hexagonal layers -- a supposition that was
finally proven in 2005 \cite{hales05,hales10}.  There is also the more
practical question of inefficiently packed objects (e.g. sand or grain
filling a container).  
At the highest packing densities that are still disordered, this
is termed random close packing, although mathematically this is
an ill-defined concept \cite{torquato00}. Experimental studies have long
recognized the influence of finite-sized and boundary effects on the packing
\cite{carman1937a,aim67,roblee58,seidler00,zhang06}. It was pointed out in the 1940's that one can extrapolate the observed packing density in containers of various sizes to infinitely large containers
\cite{verman46,brown46,scott60}; early reports noted that particles pack less efficiently in small containers.

Later work investigated how the boundaries modify the particle packing.  Particles can form layers against the wall; for
monodisperse packings, these layers can persist rather far from
the boundaries \cite{roblee58,seidler00,landry03}.  A more recent investigation
by Desmond and Weeks used simulations to study the packing of
bidisperse random close packed samples \cite{desmond09}.  They considered a 2-phase model in which packed particles are treated as a core region packed to the infinite-container volume fraction $\phi_\infty$
surrounded by boundary layer of thickness $\delta L$ packed at $\phi_\infty + \delta \phi$.  This was an extension of a
prior model that assumed $\delta L$
was the diameter of a particle.  Desmond and Weeks found evidence
that $\delta L$ was indeed of order the diameter of a particle,
and $\delta \phi < 0$:  that is, the sample packed poorly
against the boundaries.

In this paper we probe the packing of rods in small containers.
We focus on containers larger than the rod length; containers with
dimensions smaller than that would force rods to align simply to
enter the container \cite{zou96}.  Nonetheless, even in a large
container rods may align with the boundaries, resulting in a more
efficient packing. One can also imagine the opposite situation in
which boundaries prevent the rods from packing efficiently, and
perhaps support larger voids, thus decreasing the overall packing
density.  We further investigate the boundary layer properties using
x-ray tomography, confirming that the layers are thin, about half a
rod length.  The tomography results reveal that the bottom boundary
layer, top boundary layer, and side boundary layer are all
distinct, with efficient packing at the bottom layer and looser
packing at the sides and top.
Finally, we conduct simulations of confined spherocylinders in the
absence of gravity, finding boundary layers of similar thickness
to the experiments.

\section{Background}
Studies of how long thin particles pack have a long history,
including experiments using 
wooden rods \cite{zou96,milewski78,philipse96}, 
plastic rods \cite{novellani00,montillet01,stokely03,tangri17,desmond06}, 
spaghetti \cite{parkhouse95,lumay04}, 
colloidal silica particles \cite{philipse96,sacanna07}, 
glass fibers \cite{evans86}, iron rods \cite{nardin85},
lead cylinders \cite{benyahia96},
and metal wires \cite{philipse96,foumeny91,rahli99}.  Simulations have also been done with a variety of methods \cite{evans89,coelho97,williams03,abreu03,gan04,jia07,ferreiro14,zhao12rods}.
The efficiency of  such packings has relevance for nanocomposites (mixtures of polymer resin and fibers) \cite{milewski78,evans86,keledi12}, ceramics \cite{milewski78}, and
filtration and catalysis \cite{montillet01,foumeny91,coelho97}. As noted by Williams and Philipse \cite{williams03}, these studies agree fairly well on the dependence of the close packing volume fraction $\phi_{\rm rcp}$ on aspect ratio $\alpha$ ($\phi_{\rm rcp} \sim 1/\alpha$, for long rods with
$\alpha \gg 1$).  This widespread agreement suggests that
geometry plays a larger role in determining packing than the
physical properties of the particles.

Most studies used large containers to minimize the
influences of boundaries.  A few noted container size dependence:
Dixon studied boundary effects on aspect ratio 1 cylinders
(equal height and diameter), which pack less efficiently in small
containers.  For cylindrical containers of
moderate radius $R$ and large height, he found that $\phi(R)$ was
well described by a second-order polynomial in $1/R$.  Zhang {\it et
al.}~also studied the packing of aspect ratio 1 particles, using
x-ray tomography to observe the structure near the container
walls \cite{zhang06}.  A key result was that the volume fraction
within two particle diameters of the wall was lower than that of
the bulk.
Zou and Yu \cite{zou96} investigated the
packing of rods with aspect ratios 1-64 in cylindrical containers.  Their results suggested the existence of a
critical container size: for a container of large enough radius $R$
and height $H$, the results became independent of the container
size.  They also found that rods packed less densely in smaller containers.  Tangri {\it et al.} reported
\cite{tangri17} that aspect ratio 5.35 rods packed at lower $\phi$
in smaller containers.
Desmond and Franklin \cite{desmond06} observed that rods poured in smaller containing cylinders exhibit solid-body resistance to an intruder; Trepanier and Franklin \cite{trepanier10} observed similar preparation-dependent solid-body behavior in free-standing rodpiles once the confining container was removed.  Benyahia \cite{benyahia96} observed oscillations in the volume fraction near container walls for rods with aspect ratio 1, 2, and 3.

Desmond and Weeks \cite{desmond09} developed a model (building on \cite{verman46,brown46})
to understand the finite size effects of packing particles
in a container. The model assumes a boundary layer of thickness $\delta L$ with a packing fraction that differs from that found in the bulk ($\phi_\infty$) by $\delta \phi$. Applying this to a cylindrical container (radius $R$, height $H$) one can differentiate between the bulk volume
\begin{eqnarray}
V_{\rm bulk} &=& \pi(R-\delta L)^2(H-2\delta L)
\nonumber
\\
&\approx& V - \pi \delta L (2RH+2R^2)
\label{vbulk}
\end{eqnarray}
which packs at $\phi_\infty$ and the surface volume
\begin{eqnarray}
V_{\rm surf}&=& V - V_{\rm bulk} 
\nonumber
\\
&\approx& \pi \delta L (2RH+2R^2)
\label{vsurf}
\end{eqnarray}
which packs at $\phi_\infty +\delta \phi$. (Terms of
order $\delta L^2$ and higher are neglected in these definitions.)  The net volume fraction $\phi$ is a volume-weighted sum of the two regions:
\begin{equation}
\phi=\left [ V_{\rm bulk}\phi_\infty +V_{\rm surf}(\phi_\infty + \delta \phi) \right ] /V
\end{equation}
which can be shown to reduce to:
\begin{equation}
\phi(R,H) \approx {\phi_\infty} \Big[1 + 
2 \frac{\delta L}{L} \frac{\delta \phi}{\phi_\infty} 
\Big( \frac{L}{R} + \frac{L}{H} \Big) \Big].
\label{fittingold}
\end{equation}

Equation~\ref{fittingold} suggests that a plot of packing
fraction $\phi$ against a dimensionless inverse container size
$L/R+L/H$ will result in a straight line with a $y$-intercept at
the bulk/infinite-container size packing fraction $\phi_\infty$
and a slope $2\frac{\delta L}{L}\frac{\delta \phi}{\phi_\infty}$
that quantifies the impact of the boundary layer. While the slope
does not uniquely determine $\delta L/L$ or $\delta
\phi/\phi_\infty$, as we will show below x-ray tomography and
simulations do allow for independent determination of these
parameters. While Desmond and Weeks dealt with an isotropic pile
with no difference between top, bottom, and side boundaries, their model is easily generalized to account for boundary layers of differing characteristics.

\section{Procedures}

\subsection{Bulk experiments}
\label{bulkexp}

\begin{figure}[hbt]
	\begin{center}
		\includegraphics[width=4.5cm]{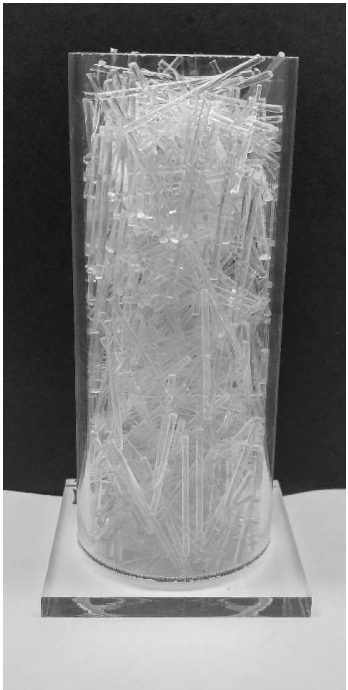}
	\end{center}
\caption{Photograph of a cylindrical container filled with aspect ratio $\alpha=20$ rods ($L=3.14$~cm).  The diameter of the container is $2R=4.49$~cm~$\approx 1.4L$ and the height is $H=12.8$~cm~$\approx 4.1L$. 
}
\label{photo}
\end{figure}

We conduct bulk experiments using two sets of containers
with different heights ($H_{\rm tall}=12.6$~cm and $H_{\rm
short}=7.4$~cm), and different diameters ranging from $2R=0.6$~cm
to 12.6~cm. We use acrylic rods of diameters $D=1/8$ inch or $1/16$
inch (0.32~cm or 0.16~cm) and hand-cut to specific lengths $L$.
The aspect ratios of the rods ($\alpha=L/D$) range from 4 to 32.
We additionally use glass spheres of radii $R=0.80$~cm as objects
of aspect ratio 1.  For each particle type, we study the packing
in various sized containers.  To avoid trivial
alignment effects in narrow containers, for each batch of rods
of length $L$ we limit the experiments to containers with $2R >
L$ \cite{zou96}.  When taking data, we gently pour the particles
into the container until the container overflows. We then remove
particles from the top of the container by hand or with tweezers so
as to minimize disturbing the particles underneath. Any particles
that extend above the top rim of the container are removed.

We then measure the total weight of the particles in the container,
and from the known particle density and known container volume, we
determine the overall volume fraction $\phi$ for that experiment.
We repeat this for five measurements per particle type per
container; below we report the mean value of the five measurements.
In nearly all cases the standard deviations are less than five
percent of the mean.  The number of measurements was too few to
examine any dependence of the standard deviation on
aspect ratio or container size.

\subsection{Tomography experiments}

\begin{figure}[b]
	\begin{center}
		\includegraphics[width=8cm]{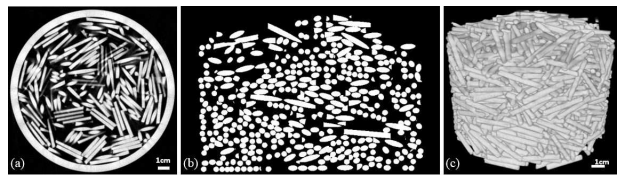}
	\end{center}
\caption{X-ray tomography data of rods with length $L=2.57$~cm,
aspect ratio $\alpha=8$ in a container with $2R=12.6$~cm~$\approx 4.9L$ and 
$H=7.4$~cm~$\approx 2.9L$.
(a) X-ray image showing a horizontal cross section through
the container.  (b) A vertical cross section, where the white
voxels are those above a threshold intensity.  The small white
dots are the corners of rods.  (c) 3D rendering of the
tomography data.
}
\label{xray}
\end{figure}

Additionally, we conduct x-ray tomography experiments to
capture the three-dimensional structural information on our rod
packings; see Fig.~\ref{xray}.  These use aspect
ratio 8 acrylic rods (0.32~cm diameter, length $L=2.57$~cm) and two cylindrical
containers, a large one with $2R=12.6$~cm~$\approx 4.9L$ and a smaller
one with $2R=4.5$~cm~$\approx 1.9L$, both with height $H=7.4$~cm~$\approx 2.9L$.
We follow the filling procedure described above, except rods are
not removed from the top.  The filled container is scanned with
a RealTime-2722 (UEG Medical Imaging
Equip. Co. Ltd.) computed tomography scanner to acquire images with 0.2~mm
resolution.  We exploit binarization and watershed segmentation
to obtain rods' centers and orientations \cite{zhang14cpb}.

We then reconstruct the whole packing to calculate the volume
fraction $\phi$.  To match the protocol used in bulk experiments (Sec.~\ref{bulkexp}) we add an additional analysis step to remove the influence of rods above the top of the container.  Rather than using the exact top of the container, we arbitrarily choose a height
near the top of the container as an artificial top container
boundary.  We remove all rods which have any portion above this
artificial boundary.  In this way, the reconstructed packing is
similar to the bulk measurements where rods were removed by hand.  We repeated the experiments five times
for each container, for a total of ten three-dimensional tomography images.

\subsection{Simulation Procedure}

We computationally generate similar packings using a technique
similar to Refs.~\cite{desmond09,ohern03}.  Particles are spherocylinders: cylinders of radius $a$ with hemispherical endcaps or, more precisely, the locus of points within a distance $a$ of a central line segment \cite{williams03}. Detection of particle contact is reduced to finding the nearest separation between line segments, as described in \cite{Pournin05}. When this separation $d$ is less than twice the radius the particles interact through a repulsive potential that is quadratic with $2a-d$. The same potential governs the interaction with container boundary.

\begin{figure}[b]
	\begin{center}
		\includegraphics[width=4.5cm]{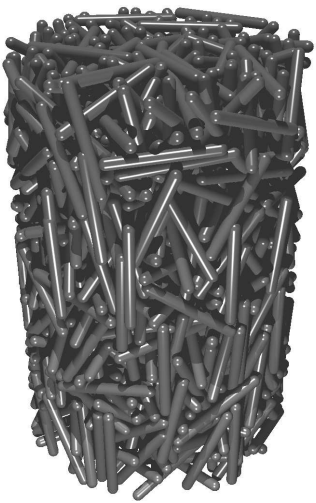}
	\end{center}
\caption{3D reconstruction of simulation data of 1024 spherocylinders with aspect ratio 9, in a container with height, radius, and spherocylinder length related by $H=3R=4.05L$.
}
\label{povray}
\end{figure}

The simulation is initialized by placing $N$ infinitesimal spherocylinders in a simulated container with random positions and orientations.  The initial length of each line segment $\ell$ is chosen to fix the aspect ratio $\alpha = (\ell+2a)/2a$. Conjugate gradient minimization is used to minimize the total potential energy as a function of the $3N$ translational + $2N$ rotational = $5N$ degrees of freedom. Once the energy is below a tolerance related to machine precision the particles are expanded, keeping the aspect ratio fixed. This increases the volume fraction and causes new contacts or overlaps, and the energy is again minimized with a conjugate gradient method. The jammed packing fraction $\phi_J$ is defined as the packing fraction at which the energy cannot be minimized to zero (or below machine precision). In practice the particles are expanded beyond $\phi_J$ and the potential energy, which increases as $(\phi - \phi_J)^\delta$, extrapolated back to find $\phi_J$.  Figure~\ref{povray} shows an example of one jammed configuration.

For each aspect ratio $\alpha$ we repeat the simulation for a
variety of initial cylindrical container sizes (height $H$, radius $R$) and number of particles $N$. Because each simulation jams at slightly different particle size, the dimensionless container size (in terms of final particle length) varies slightly. Initial parameters are chosen so that the final container sizes are comparable with the experimental systems.

\section{Results}
\subsection{Bulk packing results}

\begin{figure}[b]
	\begin{center}
		\includegraphics[width=7cm]{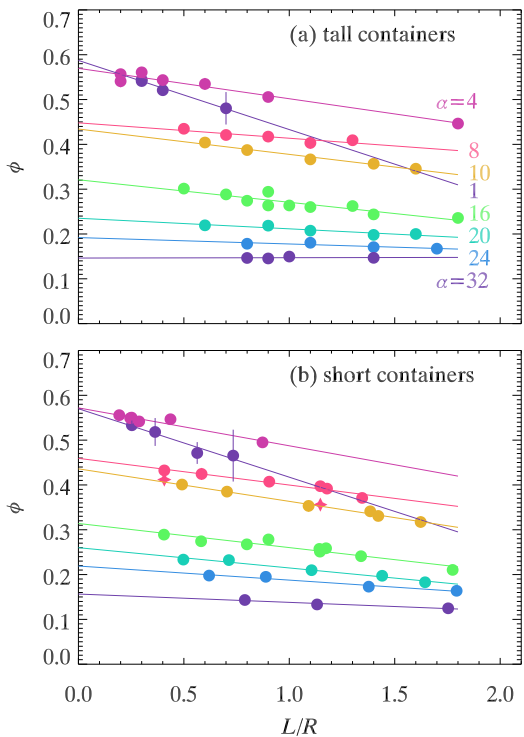}
	\end{center}
\caption{Experimentally observed $\phi$ as a function of the
inverse nondimensional container radius 
$L/R$ for (a) tall containers with $H=12.8$~cm, and (b) short containers with $H=7.4$~cm.  The data correspond to rod aspect ratio $\alpha$ as labeled in (a); aspect ratio 1 data correspond to spheres with radii $0.80$~cm.
All $\phi$ values are averages from at least five
separate trials.  The standard deviations of those trials are slightly less than the symbol size.  The lines are least squares linear fits to the data, incorporating the standard deviations as (inverse) fitting weights.  The two orange star symbols in (b) are data from the x-ray tomography experiments.
}
\label{phir}
\end{figure}

Experiments and simulations measured the bulk packing fraction for particles of aspect ratios 4-32 in containers of various sizes. For the experiment, gravity breaks the symmetry between horizontal and vertical boundaries. As a result, the simple model Eq.~\ref{fittingold} is no longer applicable and must be generalized to account for the different boundary effects.  Each of these boundaries has its own independent thickness $\delta
L/L$ and volume fraction perturbation $\delta\phi / \phi_\infty$. The derivation of Eqn.~\ref{fittingold} can be repeated, leading to a new model for $\phi(R,H)$:
\begin{equation}
\phi(R,H) \approx {\phi_\infty} \Big[1 + 
2C_R  
\Big( \frac{L}{R} \Big) + (C_T + C_B) 
\Big( \frac{L}{H} \Big) \Big]
\label{fitting}
\end{equation}
with
\begin{equation}
C_R = \frac{\delta L_R}{L} \frac{\delta\phi_R}{\phi_\infty}, 
C_T = \frac{\delta L_T}{L} \frac{\delta\phi_T}{\phi_\infty},
C_B = \frac{\delta L_B}{L} \frac{\delta\phi_B}{\phi_\infty}. 
\label{cterms}
\end{equation}
The $C$ terms describe the boundary packing effect of the radial, top, and bottom boundaries respectively. Equation \ref{fitting} suggests that the dependence of $\phi(R,H)$ on $L/R$ and $L/H$ should be investigated separately. We first plot the experimental packing fractions as functions of $L/R$, holding $H$ constant (Fig.~\ref{phir}).

Larger aspect ratios pack at lower volume fractions, as is expected from prior work
\cite{evans86,evans89,philipse96,philipse96err,williams03}.
The reasonable quality of the linear fits seen in Fig.~\ref{phir} justifies our neglect of higher order terms in the derivation of Eqns.~\ref{fittingold}, \ref{fitting}. The data in Fig.~\ref{phir} have negative slopes, indicating that particles pack to lower $\phi$ in smaller radius containers:  $C_R < 0$ in Eqn.~\ref{fitting}.

\begin{figure}[b]
	\begin{center}
		\includegraphics[width=7cm]{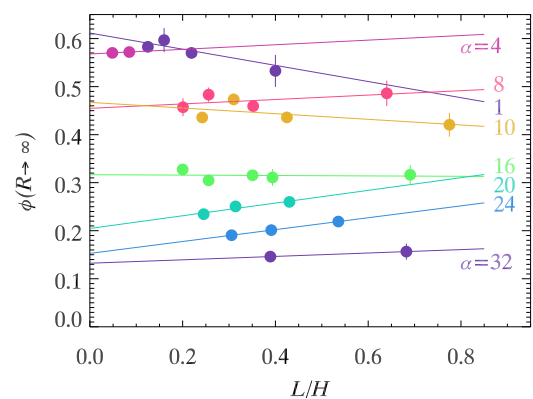}
	\end{center}
\caption{We extrapolate the data of Fig.~\ref{phir} to infinite radius containers ($L/R \rightarrow 0$) and plot the results here as a function of $L/H$.  The data are labeled by aspect ratio $\alpha$.  Additional points are from container sets with different heights (not shown in Fig.~\ref{phir}).  The error bars are due to the uncertainties arising from the linear fits to the data of Fig.~\ref{phir}.  The lines are least squares linear fits to the data, using the uncertainties as (inverse) fitting weights.
}
\label{phih}
\end{figure}

To investigate the $H$ dependence, we plot $\phi(R\rightarrow\infty,H)$  (the $y$-intercept in Fig.~\ref{phir}) against the nondimensional inverse container height ($L/H$) in Fig.~\ref{phih}. This graph has lines of both positive and negative slope, indicating that some aspect ratio rods pack worse in short containers and some pack better.  The most extreme slope is for the spherical particles, with the downward sloping data showing spheres pack better in tall containers.

\begin{figure}
	\begin{center}
		\includegraphics[width=8cm]{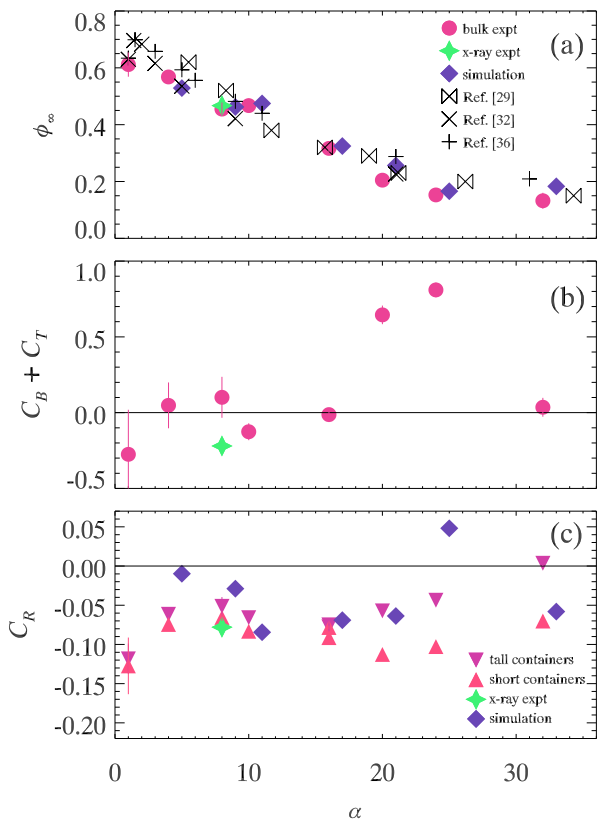}
	\end{center}
\caption{
The fitting parameters as functions of aspect ratio $\alpha$ (see Eqn.~\ref{fittingold}).
(a) $\phi_\infty$, the extrapolated volume fraction for an
infinite container.  The filled symbols are from our current work, and the other symbols are from prior simulations \cite{williams03,ferreiro14} and experiments \cite{rahli99} as indicated.  
(b) $C_B+C_T$, the combined bottom and top boundary packing parameters from the bulk data (red circles) and x-ray tomography (green star).
(c) $C_R$, the radial packing parameter.
The red downward triangles are tall container data ($H=12.8$~cm) and the red upward triangles are short container data ($H=7.4$~cm).  The green star is from x-ray data (in short containers) and the purple diamonds are from the simulation (where $C_R = C_T = C_B$).
The uncertainties for the red data in all panels are due to the uncertainties of the least squares fitting of the data of Fig.~\ref{phih} using the uncertainties of the original $\phi$ data. Uncertainties smaller than the symbol size are not shown.
}
\label{cplot}
\end{figure}

In Fig.~\ref{phih} the intercepts of the fitting lines with the vertical axis represent $\phi_\infty$, which 
describes the volume fraction of the bulk volume:  a container with infinite height and radius. We summarize the linear fit parameters in Fig.~\ref{cplot}, using the notation introduced in Eqn.~\ref{fitting}. Figure \ref{cplot}(a) shows $\phi_\infty$ as a function of the
particle aspect ratio $\alpha$.  Additionally, we
plot the $\phi_\infty$ data from prior simulations ($\times$ and
$+$, Refs.~\cite{williams03,ferreiro14}) and from prior
experiments (bowtie symbol, Ref.~\cite{rahli99}).  Our
experimental data follow a similar trend to the prior data:
large aspect ratio rods pack poorly and have a low $\phi_\infty$.
(The simulations \cite{williams03,ferreiro14,zhao12rods} found a
maximum packing for rods with $\alpha \approx 1.5$, along with
some colloidal experiments \cite{sacanna07}; we do not have any
experimental particles with this aspect ratio.)  As with prior
work \cite{stokely03}, observing the containers (Fig.~\ref{photo}) we notice trusses and tangling developing within the containers where multiple longer rods jam together and leave large voids, leading to the lower $\phi_\infty$.

Next we analyze our boundary packing parameters $C_T$, $C_B$, and $C_R$, obtained from the slopes of the lines in Figs.~\ref{phir} and \ref{phih}. The sum $C_B+C_T$ is shown in Fig.~\ref{cplot}(b), as the two parameters cannot be independently determined from the bulk experiments.  The data are scattered around $C_B+C_T = 0$, with two exceptions.  The $\alpha=1$ data point corresponds to spheres and is negative, showing that spheres pack poorly in short containers.  The $\alpha = 20,24$ data are positive, showing that these aspect ratio rods pack densely into short containers.  We know that in all cases, the top boundary layer is poorly packed due to our construction procedure (where we physically remove rods from the top of the container); thus $C_T<0$ is always true.  This means that $C_B > |C_T|$ such that their sum is positive:  the bottom layer must be quite densely packed for $\alpha=20,24$.  For the other aspect ratios where $C_B+C_T \approx 0$, it suggests that the volume fraction excess of the bottom layer is roughly equivalent to the volume fraction deficit of the top layer.

Figure \ref{cplot}(c) shows the boundary packing parameter
$C_R$ as a function of aspect ratio.  $C_R < 0$ for nearly all of the points, showing that particles pack poorly near the vertical container walls.  The two symbols are for tall containers (downward triangles) and short containers (upward triangles).
There's an interesting trend for `long rods' of $\alpha \geq 20$ 
where the tall and short container data differ.  This suggests that the corners of the container influence
the $C_R$ term for these long aspect ratio rods.  Our 
model (Eqn.~\ref{fitting}) is approximated to first order, so we lose the second-order cross-terms related to corners of height $\delta L_T$ (or $\delta L_B$) and width $\delta L_R$; it makes sense that these cross-terms could be more significant for long aspect ratio rods.  Additional data (not shown) for intermediate height containers for these long aspect ratio rods give $C_R$ results lying between the points shown in Fig.~\ref{cplot}(c), as expected.

Our data for spheres finds $C_B+C_T = -0.28,$ $C_R = -0.13$ which can be compared to $C=-0.36$ from the simulations of Desmond and Weeks \cite{desmond09}.  The results are comparable, especially given that the simulations considered a bidisperse mixture of spheres whereas our spheres are monodisperse; and the simulations did not include gravity.

We note that our results shown in Figs.~\ref{phir} and \ref{phih} differ from prior experimental work of Zou and Yu \cite{zou96}, who found a quadratic dependence of $\phi$ on $L/R$. They did see a linear dependence on $L/H$, as we do, although their data show a curious kink above a critical $H$ whence $\phi(H)$ abruptly becomes independent of $H$.  We see no evidence of a  crossover in Fig.~\ref{phih} although we explore a similar range of $L/H$.  Nor would the boundary layer model lead us to expect a kink.

Our bulk data do not allow us to distinguish the relative magnitudes
of $\delta L/L$ and $\delta \phi/\phi_\infty$. For this we turn to an x-ray tomography experiment to shed insight into how to interpret these results.

\subsection{Tomography results}

\begin{figure}
\centering
\includegraphics[width=8cm,keepaspectratio]{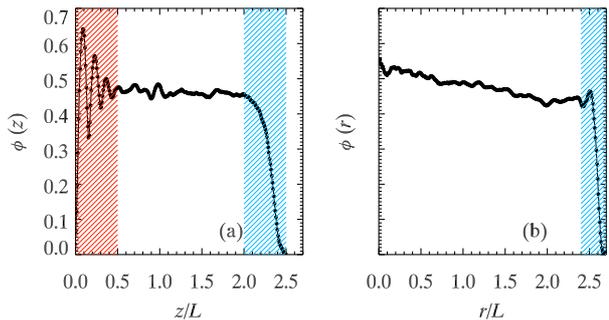}
\caption{(a) Volume fraction $\phi (z)$ as a function of distance $z$ from the bottom boundary and (b) volume fraction $\phi (r)$ as a function of distance $r$ away from the central axis.  These data correspond to a short wide container with diameter $2R \approx 4.9L$ and height $H \approx 2.9L$.  In figure (a), the red and blue shaded regions represent the top and bottom boundary layers, with boundary thickness of $\delta L=0.5L$ (estimated by eye). In figure (b), the blue shaded area represents the side boundary layer, with a smaller boundary thickness $\delta L=0.3L$ (again estimated by eye). Both $z$ and $r$ are normalized by rod's length $L$. 
}
\label{phil} 
\end{figure}

\begin{figure}
\centering
\includegraphics[width=8cm,keepaspectratio]{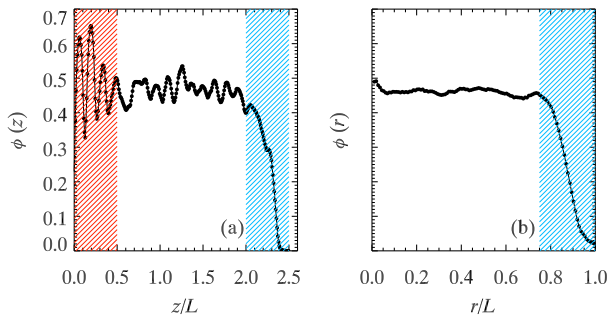}
\caption{(a) $\phi(z)$ as a function of distance $z$ away from the bottom boundary and (b) $\phi(r)$ as a function of distance $r$ away from the central axis.  These data correspond to a short narrow container with diameter $2R \approx 1.9L$ and height $H \approx 2.9L$.  All three boundary layers have the same thickness as the data shown in Fig.~\ref{phil}, as well as the same bulk value of $\phi_\infty$. 
}
\label{phis} 
\end{figure}

The x-ray tomography data sets, while limited to aspect ratio $\alpha=8$ rods, allow us to probe the volume fraction locally at every position in the container and directly look for boundary effects. In Figs.~\ref{phil}(a) and \ref{phis}(a), we plot the local volume fraction $\phi(z)$ as a function of vertical position $z$, averaged over all particles with radial positions $r<R - 0.3L$, so as to exclude from consideration any radial boundary effects.  Likewise in Figs.~\ref{phil}(b) and \ref{phis}(b) we plot $\phi(r)$ averaging over only particles with $z$ at least $L/2$ away from both the bottom and top container boundary.  Thus, in each case, $\phi(z), \phi(r) \approx \phi_\infty = 0.467 \pm 0.002$ in the middle of the container by construction.

We estimate the thickness of the top layer to again be $\delta L = L/2$, as indicated by the blue shaded regions of Figs.~\ref{phil}(a) and \ref{phis}(a).  Here the rods are quite loosely packed, with $\delta\phi / \phi_\infty = -0.39$ [Fig.~\ref{phil}(a)] and $\delta\phi / \phi_\infty = -0.48$ [Fig.~\ref{phis}(a)].  The magnitude of the reduced volume fraction make sense given our protocol for treating the top boundary of the x-ray data, where we remove rods overlapping our arbitrarily defined top boundary.  In the end, roughly half of those rods will be removed.

At the bottom of the containers [red highlighted regions of Figs.~\ref{phil}(a) and \ref{phis}(b)], there are distinct layer structures.  The spacing of these layers corresponds to the rod diameters ($L/8$ in this case), suggesting that the bottom-most layer of rods lies flat against the flat bottom boundary.  The reduction in amplitude of the fluctuations of $\phi(z)$ is due to the subsequent layers of rods lying at slight angles and packing more randomly above the flat bottom layer.  These layers are evidence of the symmetry breaking of gravity and our filling procedure.  By eye, the fluctuations appear to decay within $L/2$ from the bottom [the red shaded regions in Figs.~\ref{phil}(a) and \ref{phis}(a)], consistent with prior experiments of Montillet and Le Coq \cite{montillet01}. Interestingly, although rods in the red shaded region of Figs.~\ref{phil}(a) and \ref{phis}(a) are layered, the mean volume fraction in this region is nearly the same as the bulk region:  the fractional change in $\phi$ is only $\delta\phi / \phi_\infty=+0.01$.

X-ray tomography reveals that the side boundaries [Figs.~\ref{phil}(b) and \ref{phis}(b)] are thinner.  We estimate their thickness to be $\delta L = 0.3L$, as indicated by the shaded regions in those graphs.  Within those regions, the volume fraction is changed by $\delta \phi / \phi_\infty = -0.25$ for the larger diameter container, and -0.27 for the smaller diameter container.

\begin{figure}
    \centering
    \includegraphics[width=8cm]{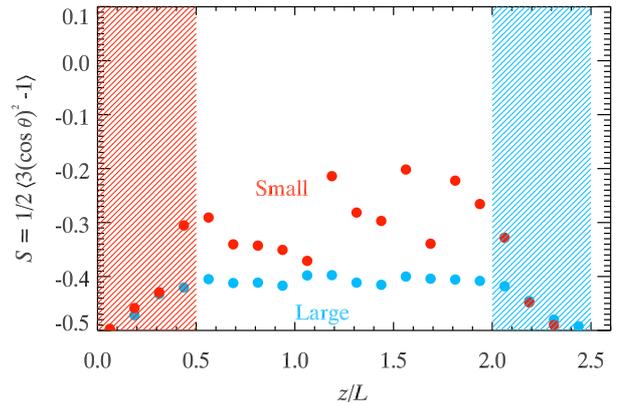}
    \caption{Nematic order parameter $S$ as a function of vertical position $z/L$ for the large ($2R = 4.9L$) and small (2R = 2.9L) containers as indicated.  $S=-1/2$ corresponds to horizontally aligned rods, and $S=0$ corresponds to isotropically oriented rods.}
    \label{nematic}
\end{figure}

Simulations, which do not incorporate gravity, show an isotropic distribution of rods. X-ray tomography allows us to study the experimental orientation distribution and, in particular, how it changes near the boundaries.  We use the nematic order parameter $S \equiv (1/2) \langle 3 \cos^2\theta - 1 \rangle$, where $\theta \in [0, \pi/2]$ is the angle with respect to the vertical axis. $S=1$ corresponds to perfectly vertically aligned rods, and $S=0$ corresponds to the case of isotropically oriented rods.  Fig.~\ref{nematic} shows the intriguing result $S<0$ for all regions in the sample, with $S \rightarrow - 1/2$ near the bottom and top of the container.  As argued above, the data of Figs.~\ref{phil}(a) and \ref{phis}(a) near $z=0$ indicate horizontally oriented rods lying in a flat layer on the bottom of the container.  For such rods, $\theta=\pi/2$ and $S = -1/2$.  At the top of the container, while the rods are not in strictly flat layers, nonetheless for a rod center to be close to $z = H$ requires the rod to be nearly horizontal, so again at the top $S \rightarrow -1/2$ makes sense.  In the middle of the container, $S$ increases toward zero but does not reach it.  This indicates that the rods are somewhat more randomly oriented, but still are influenced by gravity; they still are more horizontal than not.  Indeed, this can be directly seen in the x-ray images of Fig.~\ref{xray}(b,c), and also seen in an earlier 3D visualization experiment with aspect ratio 5 rods \cite{montillet01}.

\begin{figure}
\centering
\includegraphics[width=8cm,keepaspectratio]{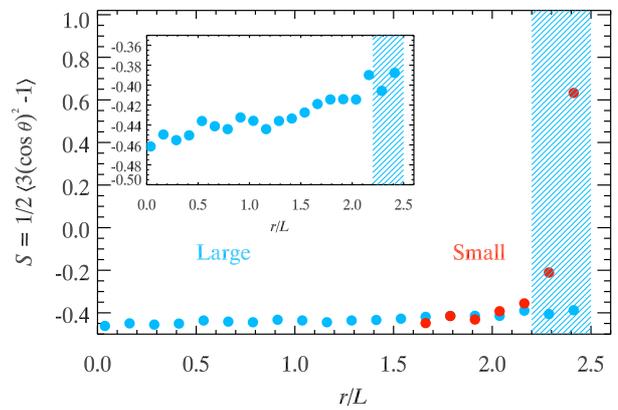}
\caption{The nematic order parameter $S$ 
as a function of the distance $r/L$ away from the center axis.  The blue shaded region represents the side boundary layer; compare with Figs.~\ref{phil}(b) and \ref{phis}(b). 
The inset graph is an expanded view of the large container data.
}
\label{ori_r} 
\end{figure}

We additionally examine $S$ as a function of radial distance $r/L$, shown in Fig.~\ref{ori_r}.  Here the smaller container data (red) are shifted horizontally so that the boundary layers overlap for the two data sets.  For the smaller container data, $S$ approaches 1 right at the container boundary, showing that the vertical container walls strongly align the rods with the vertical.  This makes sense, as the curvature of the container boundary prevents the center of a horizontal rod from coming closer than $0.09L$ to the container wall; for a rod center to be at the container wall, the rod must be vertically aligned.  This effect is diminished for the larger diameter container, for which curvature matters less; see the inset of Fig.~\ref{ori_r}.

We briefly return to Fig.~\ref{phil}(b), which shows a slight increase in $\phi(r)$ near the center of the container ($r \rightarrow 0$).  We do not know why this occurs, nor is this behavior seen in the smaller container [Fig.~\ref{phis}(b)].  However, it is consistent with the orientational behavior shown in the inset to Fig.~\ref{ori_r}, where the rods at the center of the container are more horizontal ($S$ closer to -1/2) than those farther from the center.  Presumably in the center of the container, the more horizontally aligned rods pack slightly better.

\subsection{Implications of tomography data on bulk parameters}

The $C$ terms describe the boundary packing effect of the radial, top, and bottom boundaries respectively.  Our tomography results for $\alpha=8$ can be summarized as $C_R = -0.078, C_T \approx -0.22,$ and $C_B = 0.005$.  
We can now make inferences about our model parameters based on the x-ray results.  

For much of the data of Fig.~\ref{cplot}(c), the magnitude of the product [$|C_R| = (\delta L_R / L)(|\delta \phi_R|/\phi_\infty)$] is $\sim 0.05 - 0.1$:  this suggests that either the thickness of the
boundary layer is small, or the decrease in volume fraction within
the boundary layer is small, or both.  From the x-ray data we have $\delta L_R / L \approx 1/3$ for $\alpha=8$ rods.  Assuming this is similar for other aspect ratio rods, we infer $\delta \phi/\phi_\infty \approx -0.15$ to $-0.30$ for much of the data of Fig.~\ref{cplot}(c).

For the top and bottom boundary layers, the x-ray data suggest $\delta L_T/L \approx 1/2$, $\delta \phi_T/\phi_\infty \approx -1/2$, so $C_T \approx -1/4$.  (More precisely, $C_T = -0.22$ for the $\alpha=8$ rods from the x-ray data, but that's more precision than needed at this point when considering other aspect ratio rods for which we have no direct x-ray data.)  For the cases with $C_B+C_T \approx 0$, assuming $\delta L_B / L \approx 1/2$, this then suggests $\delta \phi_B / \phi_\infty \approx +1/2$.  For $\alpha=20$ we have $C_B + C_T \approx 0.65$ and for $\alpha = 24$ we have $C_B + C_T \approx 0.80$.  Assuming the top layer has $C_T \approx -1/4$ still, it suggests these aspect ratio rods have an unusually thick or dense bottom layer.  The layering seen in Figs.~\ref{phil}(a) and \ref{phis}(a) for small $z$ has a periodicity related to the rod diameter, suggesting that the thickness of the bottom boundary layer may be more due to the rod diameter rather than rod length:  thus the thickness may be 4 rod diameters, which happens to be $L/2$ for the $\alpha=8$ rods but would be smaller for larger aspect ratio rods.  If this is true ($\delta L_B/L$ smaller for longer aspect ratio rods) then $\delta \phi_B/\phi_\infty$ could be larger than 1 for the $\alpha=20,24$ rods.  This is certainly possible as $\phi_\infty \approx 0.2$ for these rods, leaving plenty of empty volume to potentially be packed more efficiently at the bottom of the container.  Indeed, prior simulations showed that a quasi-2D layer of randomly packed rods will pack to a higher volume fraction than the same rods packed randomly in 3D due to the likelihood of 2D rods to locally align \cite{evans89}.  A nearly quasi-2D layer can pack even more densely \cite{evans89}.

\subsection{Computational results}

\begin{figure}[hbt]
\begin{center}
    \includegraphics[width=8cm]{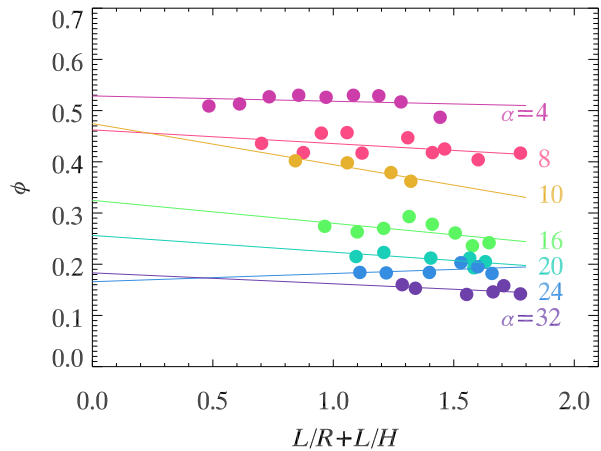}
    \end{center}
    \caption{Computational $\phi$ as a function of inverse nondimensional container size $L/R+L/H$.}
\label{phisim}
\end{figure}

Clearly the experimental results are influenced by experimental factors:  gravity-induced layering on the bottom, and an artificial particle removal procedure for the top.  To avoid these factors we turn to the simulation data.  As the simulation boundary effects should be equivalent between the cylinder top, bottom, and sides,  the simple model of Eqn.~\ref{fittingold} applies. Equivalently, we expect all three parameters in Eqn.~\ref{cterms} to be the same and the data to collapse when plotted versus the single dimensionless container size $L/R + L/H$. These data are shown in Fig.~\ref{phisim}, along with the linear fits. 

We extrapolate the linear fits to find the bulk packing fraction $\phi_\infty$ and plot this and the slope $C$ as a function of aspect ratio in Fig.~\ref{cplot}(a,c) as the violet diamond symbols.  The results are comparable to the experimental data, quantitatively for $\phi_\infty$ and the same order of magnitude for $C$ (where less agreement would be expected, given the different physics between the experiment and simulation).

%

\begin{figure}
    \begin{center}
        \includegraphics[width=7cm]{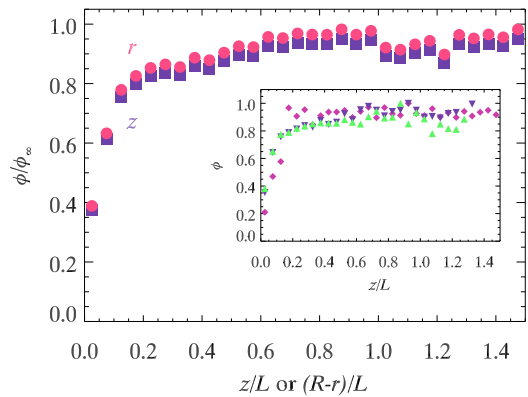}
    \end{center}

    \caption{Local packing fraction $\phi / \phi_\infty$ as a function of distance from the nearest boundary, averaged over all aspect ratios from the simulation data.  The circles are as a function of $R-r$, and the squares are as a function of $z$.  
    Inset:  Data from specific aspect ratios $\alpha=4$ (light violet diamonds), 16 (green upward triangles), and 32 (dark violet downward triangles).}
    \label{phi_bndry}
\end{figure}

Simulations allow us to investigate the boundary layer direction and confirm that, in the absence of gravity, there are no substantive differences in packing near the horizontal and vertical boundaries.  Moreover, when $\phi$ is normalized by $\phi_\infty$ and plotted as a function of $r/L$ or $z/L$, the data for different aspect ratio rods collapse.  Figure~\ref{phi_bndry} shows the data for the local packing fraction averaged over all different aspect ratios as a function of distance from the horizontal and vertical boundary. The two curves are within statistical uncertainty and reasonably well fit by a stretched exponential. Visually Fig.~\ref{phi_bndry} suggests a boundary length of about 0.25L, consistent with that seen in tomography data for the sidewalls [Figs.~\ref{phil}(b) and \ref{phis}(b)]. We note that this is independent of particle aspect ratio, suggesting a purely geometric effect.

Finally, we average over the packing fraction within the boundary length to find a local packing fraction of $0.68\phi_\infty$; thus $\delta \phi / \phi_\infty = -0.32$. Using the definition of $C$ from Eqn.~\ref{cterms} we find $C\approx -0.08$, consistent with the experimental results shown in Fig.~\ref{cplot}(b,c).

\begin{figure}
    \begin{center}
        \includegraphics[width=7cm]{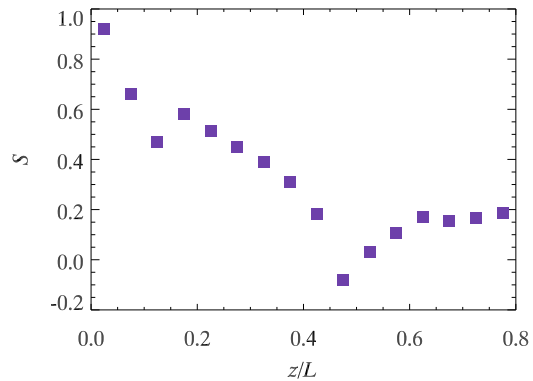}
    \end{center}

    \caption{Nematic order parameter as a function of distance from the top or bottom boundary, from the simulation data averaged over aspect ratios 4 and 12.}
    \label{sim_nem}
\end{figure}

Tomography experiments revealed the importance of gravitational-induced layering, which begins at the bottom boundary but persists throughout the bulk of the pile. In particular, the nematic order parameter $S\equiv \frac{1}{2}(3 \langle \cos^2 \theta \rangle -1)$ does not asymptote to zero as one would expect for a random distribution of particle angle (Fig.~\ref{nematic}). In simulations the situation is quite different. Fig.~\ref{sim_nem} shows the nematic order parameter as a function of distance from the nearest boundary. Rods near the boundary are aligned with boundary, as they must be, but this rapidly decays to zero. Once particles are more than half a rod-length from the side they are no longer constrained and isotropically sample the full angular space.

\section{Discussion}

It has been known that long thin particles pack poorly when packed
randomly \cite{zou96,philipse96,stokely03,evans89,williams03}.
Our results show that randomly packed long particles pack even
more poorly in smaller containers.  X-ray tomography experiments
and bulk packing data demonstrate the existence of three types
of boundary layers:  a dense bottom layer where rods lie flat,
and loosely packed side and top boundary layers.  This behavior is
captured by a generalization  of a previously derived model that
divides the space into a bulk and a boundary region, each with
their own distinct volume fractions. Our modification incorporates
the difference between bottom and top/side boundaries observed
in experiment.

The bottom layer is clearly due to gravity and our packing
protocol.  Simulations without gravitational effects reveal
more purely geometrical boundary layers of loosely packed
particles.  For both simulation and experiment, boundary layers
are roughly half a particle length thick, the most significant
finding from our work.  For all but the
most constraining containers, the impact of the boundary on
the overall volume fraction, while measurable, is generally small.

\begin{acknowledgements}

The work of J.O.F. and E.R.W was supported by grants from the
National Science Foundation (CMMI-1250199/1250235, DMR-1609763). The work of S.V.F. and S.F.P. was supported by grants from the National Science Foundation (CBET-1133722, CBET-1438077).

\end{acknowledgements}


\end{document}